\documentclass[11pt,a4paper]{article}
\pdfoutput=1

\usepackage{jcappub}

\newcommand{\bea}{\begin{eqnarray}}
\newcommand{\eea}{\end{eqnarray}}
\newcommand{\beq}{\begin{equation}}
\newcommand{\eeq}{\end{equation}}

\title{Inflection point inflation: WMAP constraints and a solution to the fine tuning problem}

\author[a]{Shaun Hotchkiss,}
\author[b]{Anupam Mazumdar}
\author[c]{and Seshadri Nadathur}

\affiliation[a]{Department of Physics, University of Helsinki and Helsinki Institute of Physics, P.O. Box 64, FIN-00014 University of Helsinki, Finland}
\affiliation[b]{Physics Department, Lancaster University, Lancaster LA1 4YB, UK\\
Niels Bohr Institute, Copenhagen, Blegdamsvej-17, Denmark}
\affiliation[c]{Rudolf Peierls Centre for Theoretical Physics, University of Oxford, Oxford OX1 3NP, UK}

\emailAdd{shaun.hotchkiss@helsinki.fi}
\emailAdd{a.mazumdar@lancaster.ac.uk}
\emailAdd{seshadri@thphys.ox.ac.uk}

\abstract{We consider observational constraints and fine-tuning issues in a renormalizable model of inflection point inflation, with two independent parameters. We derive constraints on the parameter space of this model arising from the WMAP 7-year power spectrum. It has previously been shown that it is possible to successfully embed this potential in the MSSM. Unfortunately, to do this requires severe fine-tuning. We address this issue by introducing a hybrid field to dynamically uplift the potential with a subsequent smooth phase transition to end inflation at the necessary point. Large parameter regions exist where this drastically reduces the fine-tuning required without ruining the viability of the model. A side effect of this mechanism is that it increases the width of the slow-roll region of the potential, thus also alleviating the problem of the fine-tuning of initial conditions. The MSSM embedding we study has been previously shown to be able to explain the smallness of the neutrino masses. The hybrid transition does not spoil this feature as there exist parameter regions where the fine-tuning parameter is as large as $10^{-1}$ and the neutrino masses remain small.}


\arxivnumber{1101.6046}

\begin{document}
\maketitle
\flushbottom

\section{Introduction}
\label{section:introduction}

An outstanding goal for the theory of primordial inflation is to connect it to particle physics, and in particular to the Standard Model (SM) and its extensions \cite{RM}. For this to happen it would appear that a low inflationary scale and sub-Planckian VEVs are necessary. These features are naturally present when inflation is generated about a {\it point of inflection} in the potential, due to the flatness of the potential at the inflection point. This idea can be illustrated with a simple renormalizable scalar potential with two independent parameters, $A$ and $B$:
\beq
\label{genericpotential}
 V(\phi)=A\phi^2-C\phi^3+B\phi^4\,,
\eeq
where $C$ is determined in terms of $A$ and $B$ in order to obtain a point of inflection suitable for inflation. The VEV at which inflation occurs is closely related to the two independent parameters and can take a wide range of values below $M_{P}=2.4\times 10^{18}$~GeV for different values of $(A,B)$. 

The above renormalizable potential is sufficiently simple that it can be embedded in a range of particle theories beyond the SM. Of particular interest is a model of low scale supersymmetry (SUSY), where the origin of $\phi$ can be directly linked to SUSY partners of the SM Higgs and leptons, along with the right-handed (RH) neutrinos \cite{Allahverdi:2006cx,Allahverdi:2007wt}.\footnote{The combination is a {\it gauge invariant} D-flat direction of $MSSM\times U(1)_{B-L}$. (For a review on SUSY flat directions, see~\cite{MSSM-REV}).} This model has been shown to produce a power spectrum of perturbations that is consistent with observation \cite{Allahverdi:2007wt}, while explaining the small scale of the observed neutrino masses, and providing a dark matter candidate from the RH sneutrino component of the inflaton \cite{Allahverdi:2007wt} in a simple extension of minimal supersymmetric Standard Model (MSSM).\footnote{The first examples of an MSSM {\it gauge invariant} inflaton are given in \cite{Allahverdi:2006iq,Allahverdi:2006we} and the parameter space for the detection of MSSM inflatons and neutralino dark matter at the LHC was studied in \cite{Allahverdi:2007vy,Allahverdi:2010zp}.}

The advantage of such a particle physics embedding is that the model parameters are motivated by low scale SUSY within a \emph{visible} sector. Therefore, one can track the thermal history of the universe and probe the inflaton origin at the LHC, while also constraining the potential from cosmic microwave background (CMB) observations. In Section~\ref{section:constraints} of this paper we investigate which regions of parameter space are consistent with constraints from the WMAP 7-year results.

It is known that in order to maintain sufficient flatness of the potential to reproduce the observed CMB spectrum, a fine-tuning of the parameters $A$, $B$ and $C$ is required~\cite{Allahverdi:2006we,Bueno Sanchez:2006xk,Lalak:2007rsa,Enqvist:2007tf}. At low scales this tuning is very acute and a significant challenge to overcome. For high or intermediate scale inflation the required tuning can be reduced to some extent but remains problematic. Although the string landscape can perhaps naturally account for the fine tuning of soft SUSY breaking terms from degenerate vacua~\cite{Allahverdi:2007wh}, a dynamical solution to the problem is desirable.\footnote{In the context of a non-renormalizable potential for inflection point inflation, a different approach to the fine-tuning problem has been presented in ref.~\cite{Allahverdi:2010zp}. Using the renormalization group equations, the tuning of the ratio of SUSY breaking terms can instead be viewed as an equal tuning of the non-renormalizable coupling.}

The fine-tuning required to fit the spectrum constraints can also be reduced by simply raising the scale of inflation, as first pointed out in \cite{Enqvist:2010vd}. However, on its own this violates the requirement to generate a suitable number of e-folds. We propose to include a new hybrid scalar field which provides a vacuum energy while it remains trapped in a false minimum. On being released from this false minimum, the field rolls quickly to its true minimum and brings slow-roll to a premature end, in exactly the same manner as hybrid inflation~\cite{Linde:1993cn}. This extension can significantly reduce the amount of fine-tuning required in the model, while still matching observations and the e-fold constraint, without ruining any of the attractive features of the MSSM embedding.

In Section~\ref{section:MSSM} we briefly discuss the aspects of the SUSY embedding of \eqref{genericpotential}, define the measure of fine-tuning of the potential, and obtain expressions for the slow-roll parameters that are used later in the paper. The bulk of Section~\ref{section:MSSM} is however devoted to a calculation of the e-fold number that corresponds to the observed CMB scales. This calculation is not original but it provides important results that are used in Sections~\ref{section:constraints} and~\ref{section:hybrid}, which contain the new results of the paper. In Section~\ref{section:constraints} we investigate the region of parameter space that allows for a period of inflation that is consistent with the e-fold constraint and constraints from the WMAP 7-year power spectrum \cite{Komatsu:2010fb}. In Section~\ref{section:hybrid}, we introduce the hybrid extension to the model and show how it reduces the required fine-tuning.

\section{Slow-roll parameters and the e-folding number}
\label{section:MSSM}

Let us consider a re-parameterisation of eq.~(\ref{genericpotential}):
\beq
\label{potential}
V\left(\vert\phi\vert\right)=\frac{m_\phi^2}{2}\vert\phi\vert^2+\frac{h^2}{12}\vert\phi\vert^4-\frac{Ah}{6\sqrt{3}}\vert\phi\vert^3\,,
\eeq
The potential will have a region suitable for inflation if the mass term satisfies the condition
$A \approx 4 m_\phi $. This form of the potential was first motivated in Refs.~\cite{Allahverdi:2006cx,Allahverdi:2007wt} from a low scale extension of the MSSM with an additional $U(1)_\mathrm{B-L}$, which can tie the light neutrino masses to the flatness of the inflaton potential. Here $\phi$ contains the RH sneutrino field $\tilde N$, Higgs $H_{u}$, and the left handed slepton field $\tilde L$. The potential can be derived from the  superpotential term
\beq
\label{superpotential}
W\supset h\mathbf{NH_uL}\,,
\eeq
where $\mathbf{N}$, $\mathbf{L}$ and $\mathbf{H_u}$ are superfields, and $h$ is the Yukawa coupling. In this particular case the mass of the inflaton will be given by:
\beq
\label{mphi}
m_\phi^2=\frac{m_{\tilde{N}}^2+m_{H_u}^2+m_{\tilde{L}}^2}{3} .
\eeq
Therefore, the mass of the Higgs field is tied to the mass of the inflaton. If $h\sim 10^{-12}$, it is possible to also tie the smallness of the observed neutrino masses to the flatness of the inflaton potential \cite{Allahverdi:2006cx,Allahverdi:2007wt}. We stress these points because their implications are re-considered in sections~\ref{section:constraints} and \ref{section:hybrid}.

Following convention, we parameterize the fine-tuning of the potential as
\beq
\delta \equiv A^2/(16m_\phi^2) \equiv 1-\beta^2/4\, ,
\eeq
where $\beta$ may be either real or imaginary, corresponding to $\delta<1$ or $\delta>1$ respectively. The ratio of SUSY breaking terms $A/4m_\phi$ must be tuned to accuracy $\mathcal{O}(\vert\beta\vert^2)\ll1$. Under this condition, the potential has a point of inflection at $\phi=\phi_0$ such that $V^{\prime\prime}(\phi_0)=0$, where the $\prime$ denotes differentiation with respect to $\phi$. The point of inflection is\footnote{In fact there are two solutions for the inflection point $\phi_0$ but we shall only consider the larger one. Our analysis is valid as long as the the field is always in the vicinity of this point.} 
$
\phi_0 = \sqrt{3}m_\phi/h
$.
In the vicinity of $\phi_0$ the potential can be written as the truncated Taylor expansion:
\beq
\label{Taylorexp}
V(\phi)=V_0+\alpha(\phi-\phi_0)+\frac{\gamma}{6}(\phi-\phi_0)^3\,,
\eeq
where $V_0\equiv V(\phi_0)$, $\alpha\equiv V^\prime(\phi_0)$ and $\gamma\equiv V^{\prime\prime\prime}(\phi_0)$. This expansion will  be valid provided that
\beq
\label{Taylorcondition1}
\vert\alpha\vert\gg\left\vert\frac{d^mV}{d\phi^m}(\phi_0)\right\vert\vert\phi_e-\phi_0\vert^{m-1}
\eeq
and
\beq
\label{Taylorcondition2}
\vert\gamma\vert\gg\left\vert\frac{d^mV}{d\phi^m}(\phi_0)\right\vert\vert\phi_e-\phi_0\vert^{m-3}
\eeq
for $m\ge4$, where $\phi_e$ is the value of the field at the end of inflation. 
The terms in this expansion are~\cite{Allahverdi:2006we} $V_0\propto m_\phi^2\phi_0^2\propto m_\phi^4/h^2$; $\gamma\propto m_\phi^2/\phi_0\propto m_\phi h$ and
\beq
\label{saddlealpha}
\alpha = \frac{\sqrt{3}m_\phi^2 }{4h}\beta^2 \,  +\mathcal{O}(\beta^4)\, .
\eeq
Note that the Hubble rate during inflation is $H_\mathrm{inf}\approx V_0^{1/2}/\sqrt{3}M_P$.

From the form of the potential in eq.~\eqref{Taylorexp} we may write the slow-roll parameters $\epsilon\equiv (M_P^2/2)(V'/V)^2$ and $\eta\equiv M_p (V''/V)$ explicitly as:
\bea
\label{eps}
\epsilon(\phi)&=&\frac{M_P^2}{2 V_0 ^2} \left(\alpha+\frac{\gamma}{2}\left(\phi-\phi_0\right)^2 \right)^2 \\
\label{eta}
\eta(\phi)&=&-\frac{\gamma M_P^2}{V_0} \left(\phi_0-\phi\right) .
\eea
If inflation ends at field value $\phi=\phi_e$, the number of e-folds of inflation produced as the field rolls from $\phi$ to $\phi_e$ is given by
\beq
\label{Nactual}
\mathcal{N\left(\phi\right)}=\int_{\phi}^{\phi_e}\frac{H d\phi}{\dot\phi}=\frac{V_0}{M_P^2}\sqrt{\frac{2}{\alpha \gamma}}\left[F\left(\phi_e \right) - F\left(\phi\right)\right]
\eeq
where $F(z) \equiv \mathrm{arccot}\left(\sqrt{\frac{\gamma}{2\alpha}}(z-\phi_0)\right)$. When $\beta$ is imaginary ($\alpha<0$) the corresponding expression is found by analytic continuation. While the Taylor expansion \eqref{Taylorexp} is not strictly necessary to calculate $\mathcal{N}$ it does allow for this closed-form, analytic expression for $\mathcal{N}$. Calculating $\mathcal{N}$ numerically gives equivalent results but significantly lengthens the computation time. This is particularly true of the calculation in Section~\ref{section:hybrid}, where \eqref{Nactual} must be inverted. This would make the production of figures \ref{figure:m-h} and \ref{figure:Vc-h10} unfeasible. 

Slow-roll ends at the field value $\phi_{e}$ at which $\vert\eta\vert\sim1$. This is determined by equation~\eqref{eta}:
\beq
\label{phieta}
\phi_e\sim\phi_0-\frac{V_0}{\gamma M_P^2}.
\eeq
An upper bound on the maximum number of e-foldings between the time when the observationally relevant perturbations were generated and the end of inflation can be derived as in Ref.~\cite{Liddle:2003as}, under the assumption that the energy scale of inflation is roughly constant during inflation (which is valid as $\epsilon\ll |\eta|\ll 1$). Using the WMAP pivot scale $k_\mathrm{pivot}=0.002~\mathrm{Mpc}^{-1}$ and the current best-fit values of the cosmological parameters, we find this to be\footnote{WMAP quote observational results at the pivot scale, $k_\mathrm{pivot}$, which is not equal to the horizon scale $a_0H_0$. The difference between these two scales corresponds to $\sim 2$ e-folds.}
\beq
\label{Nmax}
\mathcal{N}\leq\mathcal{N}_{\mathrm{pivot}}\equiv64.7+\ln\left(\frac{V_0^{1/4}}{M_P}\right) .
\eeq
Note that $\mathcal{N}_\mathrm{pivot}$ is dependent on the scale of inflation and therefore on the parameters $m_\phi$ and $h$.

In fact, if the origin of the inflaton is given by eq.~\eqref{superpotential}, the time scale for transferring the inflaton energy into relativistic species  is negligible compared to the Hubble time \cite{Allahverdi:2006we} so we can treat this process to be instantaneous and the inequality in eq.~\eqref{Nmax} is saturated.\footnote{Note that to determine the necessary number of e-folds we only need to know the point at which the equation of state changes from inflaton- to radiation- or matter-domination~\cite{Burgess:2005sb}. The time scale for the thermalization of all the relativistic species could be different.} In order to account for errors in the value of $\mathcal{N_\mathrm{pivot}}$ we can take the constraint to be $\mathcal{N}\simeq\mathcal{N_\mathrm{pivot}}$. In fact, the error introduced due to uncertainties in the cosmological parameter values, the slow-roll approximation and the assumption of instantaneous transition from radiation- to matter-domination of the universe is very small, of order at most one e-fold. We can therefore safely require the e-fold constraint to be satisfied within $\pm5$ e-folds. As shown in the next section, even this conservative assumption imposes tight constraints on the allowed values of $m_\phi$ and $h$.


\section{WMAP constraints on $(m_{\phi},h)$ and fine-tuning of parameters}
\label{section:constraints}

\begin{figure*}[tbp]
\begin{center}
\includegraphics[scale=0.6]{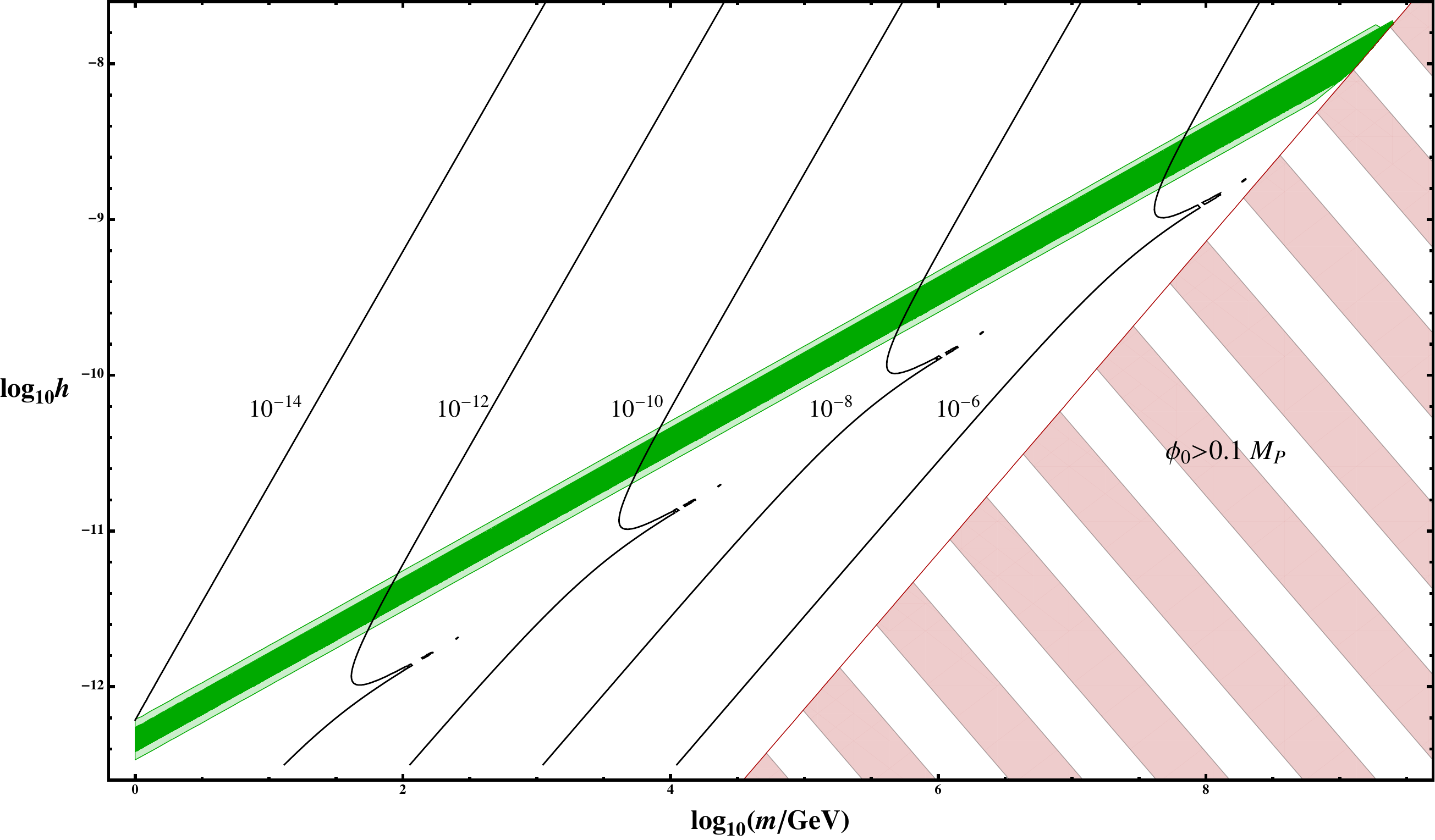}
\caption{Regions of parameter space for the potential in eq.~\eqref{potential} that satisfy the WMAP 7-year  constraints on the amplitude and spectral index of the power spectrum and also match the e-fold constraint. We minimize $\vert\mathcal{N}-\mathcal{N}_\mathrm{pivot}\vert$ over the range of $\beta$ allowed by the 95\% C.L. constraints on $\mathcal{P}_\mathrm{R}$ and $n_s$. The dark green central contour shows the region for which $\left(\vert\mathcal{N}-\mathcal{N}_\mathrm{pivot}\vert\right)_\mathrm{min}\leq1$, the light green contour is for $\left(\vert\mathcal{N}-\mathcal{N}_\mathrm{pivot}\vert\right)_\mathrm{min}\leq5$. Contour lines of $\vert\beta\vert$ are shown in black, for the values of $\vert\beta\vert$ indicated. The red striped region is excluded by the requirement that the inflection point be much less than the Planck scale.}
\label{figure:m-h}
\end{center}
\end{figure*}

If the perturbations relevant to the CMB spectrum observed today were generated at a field value $\phi=\phi_\mathrm{CMB}$, the amplitude of the power spectrum and the scalar spectral index are given by:
\bea
\label{Pr}
\mathcal{P}_R&=& \frac{1}{24\pi^2M_P^4}\frac{V_0}{\epsilon\left(\phi_\mathrm{CMB}\right) } \\
n_s &=& 1+2\eta\left(\phi_\mathrm{CMB}\right) - 6\epsilon\left(\phi_\mathrm{CMB}\right) .
\eea
As $\epsilon\ll\vert\eta\vert$ we can approximate the spectral index as 
\beq
\label{ns}
n_s = 1+2\eta\left(\phi_\mathrm{CMB}\right).
\eeq
The latest data from the WMAP 7-year release suggest a power spectrum with $\mathcal{P}_R=(2.43\pm0.11)\times10^{-9}$ and a spectral index of $n_s=0.967\pm0.014$ for models with no \textquoteleft running' of the spectral index \cite{Komatsu:2010fb}. (In all the cases considered in this paper, the running is found to be negligible, so we do not mention it explicitly.) Given these constraints eqs.~\eqref{Pr} and \eqref{ns} may be inverted to obtain the values $\epsilon_\mathrm{best}$ and $\eta_\mathrm{best}$ that will produce the required best-fit power spectrum, and the range of $\epsilon$ and $\eta$ values that lie within the $95\%$ C.L. The field value $\phi_\mathrm{CMB}$ at which the perturbations are generated is then chosen to be
\beq
\label{phiCMB}
\phi_\mathrm{CMB}=\phi_0-\frac{V_0}{\gamma M_P^2}\vert\eta_\mathrm{best}\vert\,.
\eeq
Equation~\eqref{eps} then requires that the first derivative of the potential satisfy
\beq
\label{alpha}
\alpha_\mathrm{CMB}=\frac{\sqrt{2\epsilon_\mathrm{best}}V_0}{M_P}-\frac{V_0^2\eta_\mathrm{best}^2}{2\gamma M_P^4}\,,
\eeq
from which we obtain the required value of the fine-tuning parameter $\beta_\mathrm{CMB}$ using eq.~\eqref{saddlealpha}. 

Equation~\eqref{alpha} is the condition for a given potential of the form of \eqref{Taylorexp} to have a region in which perturbations with the best-fit power spectrum can be produced. Equation~\eqref{phiCMB} identifies this region. For any combination of parameters $m_\phi$ and $h$ it is always possible to make this choice, but the value of $\beta_\mathrm{CMB}$ thus obtained may not simultaneously satisfy $\mathcal{N}\simeq\mathcal{N}_\mathrm{pivot}$, with  $\mathcal{N}$ calculated using equations \eqref{Nactual}, \eqref{phieta}, \eqref{phiCMB} and \eqref{alpha}. However, one has the freedom to choose a range of values for $\beta$ around $\beta_\mathrm{CMB}$, governed by the acceptable range in the slow-roll parameters. If for any value of $\beta$ in this range $\mathcal{N}\simeq\mathcal{N}_\mathrm{pivot}$ to within the desired uncertainty then that combination of $m_\phi$ and $h $ can be consistent with current WMAP limits.

In figure~\ref{figure:m-h} we show the allowed range of parameters $m_\phi$ and $h$. We impose the conservative constraint that the inflaton VEV should be sub-Planckian, $\phi_0<0.1M_P$, and require that the conditions in eqs.~\eqref{Taylorcondition1} and \eqref{Taylorcondition2} be satisfied. Note that if $\beta$ is imaginary the potential develops a false minimum separated from the true minimum by a barrier. Near the peak of this barrier a self-reproduction regime exists where quantum diffusion dominates over the classical evolution \cite{Allahverdi:2006we}. We have also checked that none of the allowed regions in figure~\ref{figure:m-h} are in this regime. Two different contours are shown, using the criteria $\vert\mathcal{N}-\mathcal{N}_\mathrm{pivot}\vert_\mathrm{min}\leq5$ and $\vert\mathcal{N}-\mathcal{N}_\mathrm{pivot}\vert_\mathrm{min}\leq1$ over the allowed range of $\beta$ to define consistency with the WMAP data.\footnote{These results may be compared with those obtained in Ref.~\cite{Allahverdi:2006iq, Bueno Sanchez:2006xk}. The small difference probably arises because the authors of these papers have taken $\mathcal{N}=50$ throughout the range of the plot, without accounting for changes in $\mathcal{N}_\mathrm{pivot}$ due to the change in the energy scale of inflation.} Also plotted are contour lines of $\vert\beta_\mathrm{CMB}\vert$ --- the kink in the contours indicates the transition from imaginary to real $\beta_\mathrm{CMB}$. 

Note that for \emph{all} $\left(m_\phi,h\right)$ coordinates in figure~\ref{figure:m-h} it is possible to generate perturbations with the desired power spectrum \emph{on some scales}. However for coordinates lying below the contours shown, these scales are too large and have not re-entered our horizon yet, i.e. they correspond to $\mathcal{N}>\mathcal{N}_\mathrm{pivot}$. The opposite is true for the region above the allowed contour, where not enough e-folds are generated. This region could be accessible if the period of reheating can be delayed by several e-folds~\cite{Allahverdi:2008pf,Enqvist:2010ky}.

\subsection{Low scale inflation and fine-tuning}

It can be seen that for low inflaton masses, the coupling $h$ is required to be very small for the model to simultaneously match the spectrum and e-fold constraints. Such a low value of $h$ can in fact provide an explanation for the low masses of the neutrinos if the origin of the inflaton is given by eq.~\eqref{superpotential}. If the neutrino is of Dirac type its mass is given by $m_\nu = h\left<H_u\right>$ where $\left<H_u\right>\simeq174$ GeV is the Higgs VEV. For $m_\phi\sim10$ GeV, $h\sim10^{-12}$, this gives $m_\nu\sim0.1$ eV. This is in line with current constraints from cosmology and atmospheric neutrino oscillations as detected by the Super-Kamiokande experiment. Note that the inflaton mass, eq.~(\ref{mphi}), is constrained to be small at the scale of inflation when the VEV is $~\phi_{0}\sim 10^{14}$~GeV, during which the relevant perturbations are created. However the inflaton mass evolves with the VEV (and the energy scale) due to the SM gauge interactions, and at the LHC scale is much higher~\cite{Allahverdi:2007wt}.

The fine-tuning required for such low scale inflation is very acute: note that for inflaton mass around the TeV scale, we have $\beta_\mathrm{CMB}\lesssim10^{-10}$, which corresponds to a severe tuning of the ratio of the SUSY breaking terms.

\subsection{High and intermediate scale inflation}

The inflaton mass can vary over a wide range of scales, subject to an absolute upper bound of $m_\phi\sim\mathcal{O}(10^9)$ GeV obtained from the condition that the VEV during inflation be much less than the Planck scale.

For the model in eq.~\eqref{superpotential} it can be seen that at large $m_\phi$ the Yukawa coupling $h$ becomes much larger than the upper bound $h\sim10^{-12}$ for the Dirac mass of the neutrino to not exceed current constraints. If the neutrino is not of Dirac type this does not impose a constraint on the model. In fact the RH neutrinos can obtain a Majorana mass $M$ through the breaking of the $U(1)_\mathrm{B-L}$ symmetry, which gives the active neutrino masses by the seesaw relation $h^2\left<H_u\right>^2/M$ \cite{Mohapatra:2005wg,deGouvea:2005er}. For reasonable values of $M\gtrsim\mathcal{O}(1~\mathrm{TeV})$ the active neutrino mass remains $h^2\left<H_u\right>^2/M\ll0.1$ eV, which is acceptable for the lightest neutrino species.

Raising the scale of inflation can somewhat reduce the fine-tuning, but it can be seen that the largest value of $\beta_\mathrm{CMB}$ is bounded by:
\beq
\beta_\mathrm{CMB} \lesssim 10^{-6}\,.
\eeq
This still represents a serious problem. We address how to improve on this in the next section.

\section{A dynamical solution to the fine tuning problem}
\label{section:hybrid}

In this section, we will take the origin of the inflaton to be as in eq.~\eqref{superpotential}. The fine-tuning of soft SUSY-breaking parameters is required in order make the potential sufficiently flat. Raising the potential during inflation increases the flatness of the potential, thus ameliorating the tuning \cite{Enqvist:2010vd}. However, this also leads to too many e-folds of inflation, $\mathcal{N}>\mathcal{N}_\mathrm{pivot}$ and thus cannot by itself provide a consistent solution. When viewed from this perspective, it is clear that both problems can be solved simultaneously by introducing a hybrid mechanism~\cite{Linde:1993cn} to bring a premature end to inflation.

We therefore introduce a new scalar field, $S$, which acquires a VEV and couples to the inflaton, thus altering the basic potential in eq.~\eqref{potential} to the form
\bea
\label{hybridpotential}
V(\vert\phi\vert,S) =&&\frac{1}{2}m^2|\phi|^2 + \frac{h^2}{12}\vert\phi\vert^4 - \frac{Ah}{6\sqrt{3}}\vert\phi\vert^3  \nonumber \\
&&+ \left(S^2 - V_c^{1/2}\right)^2 + \frac{1}{2}g^2\vert\phi\vert^2S^2\,.
\eea
During slow-roll the $S$ field obtains a large effective mass from the VEV of $\phi$, i.e. $m_{\mathrm{eff},S}\gg H_\mathrm{inf}$, and rolls quickly within one Hubble time to get
trapped at its local minimum $S\approx0$.  The effect on the potential is to add a constant term $V_c$ which flattens the potential and reduces the required fine-tuning in $\beta$.  The Taylor expansion of eq.~\eqref{Taylorexp} is the same, except for the modification
\beq
\label{modifiedV0}
V_0\rightarrow \tilde{V}_0=V_c + \frac{m_\phi^4} {4h^2} + \mathcal{O}(\beta^2)\,,
\eeq 
and provided that the conditions in eqs.~\eqref{Taylorcondition1} and \eqref{Taylorcondition2} are satisfied, the analysis of section~\ref{section:constraints} remains unchanged, except that an extra degree of freedom is now introduced via $V_c$. This situation persists until $\phi$ rolls to a critical value $\phi_c$ such that
 \beq
 \label{newphie}
 \phi_c = \frac{2V_c^{1/4}}{g}\,,
 \eeq
at which point $S$ is released from the origin and rolls to its global minimum, acquiring a VEV $\left<S\right>=V_c^{1/4}$. Provided that $V_c >V_0$, this leads to a sudden steepening of the potential and brings an end to slow-roll inflation. For reasonable $V_c$, $\eta_S\equiv m_S^2M_P^2/V_c\gg1$ where $m_S^2/2=2V_c^{1/2}$ is the negative mass term for $S$, and so inflation will end promptly~\cite{Copeland:1994vg}. (Models with more complicated behaviour after the hybrid transition \cite{Clesse, Kodama} are not relevant in this context.) The VEV acquired by the $S$ field gives an effective mass to the inflaton, $m^{2}_\phi\rightarrow\tilde{m}^{2}_\phi=m^{2}_\phi+g^{2}V_c^{1/2}$.

From eqs.~\eqref{alpha} and \eqref{saddlealpha} we see that for any given values of the parameters $\left(m_\phi,h\right)$,  increasing $V_c$, and thus $\tilde{V}_0$, increases $\vert\beta\vert$ and alleviates the fine-tuning of the potential. Clearly there is a threshold above which increasing $V_c$ requires $\beta$ to be imaginary. We check that the field is never required to be in the self-reproduction region of the potential, therefore this does not pose a problem. However, $V_c$ cannot be increased without limit, as with increasing $\vert\beta\vert$ there will come a point when the condition in eq.~\eqref{Taylorcondition1} is no longer satisfied and our inflection point analysis will no longer be valid.

For any given parameter combination $\left(m_\phi,h\right)$, after having chosen some value of $V_c$ and the corresponding $\beta_\mathrm{CMB}$ which will produce exactly the best-fit WMAP power spectrum amplitude and spectral index, we can invert eq.~\eqref{Nactual} with $\mathcal{N}$ set equal to $\mathcal{N}_\mathrm{pivot}$ to find the appropriate field value $\phi_e$ for the end of inflation such that scale at which the perturbations with this power spectrum are produced should be exactly the scale observed by WMAP today. Setting the critical value $\phi_c=\phi_e$ we can then find the coupling strength $g$ for which the hybrid scalar field will bring inflation to an end at exactly this field value.  In this way we can choose the parameters $V_c$ and $g$ appearing in the potential in eq.~\eqref{hybridpotential} while still ensuring that the constraint on the number of e-folds of inflation since the generation of the observed perturbations is still satisfied. This gives us the power to increase the value of $\beta$ and thus ameliorate the fine-tuning problem. 

In this analysis for simplicity we disregard any uncertainty in the power spectrum amplitude and spectral index, and in $\mathcal{N}_\mathrm{pivot}$. Accounting for these will provide a range of acceptable values of the coupling $g$.

\subsection{WMAP constraints and solution to the fine tuning problem}

\begin{figure*}[tbp]
\begin{center}
\includegraphics[scale=0.6]{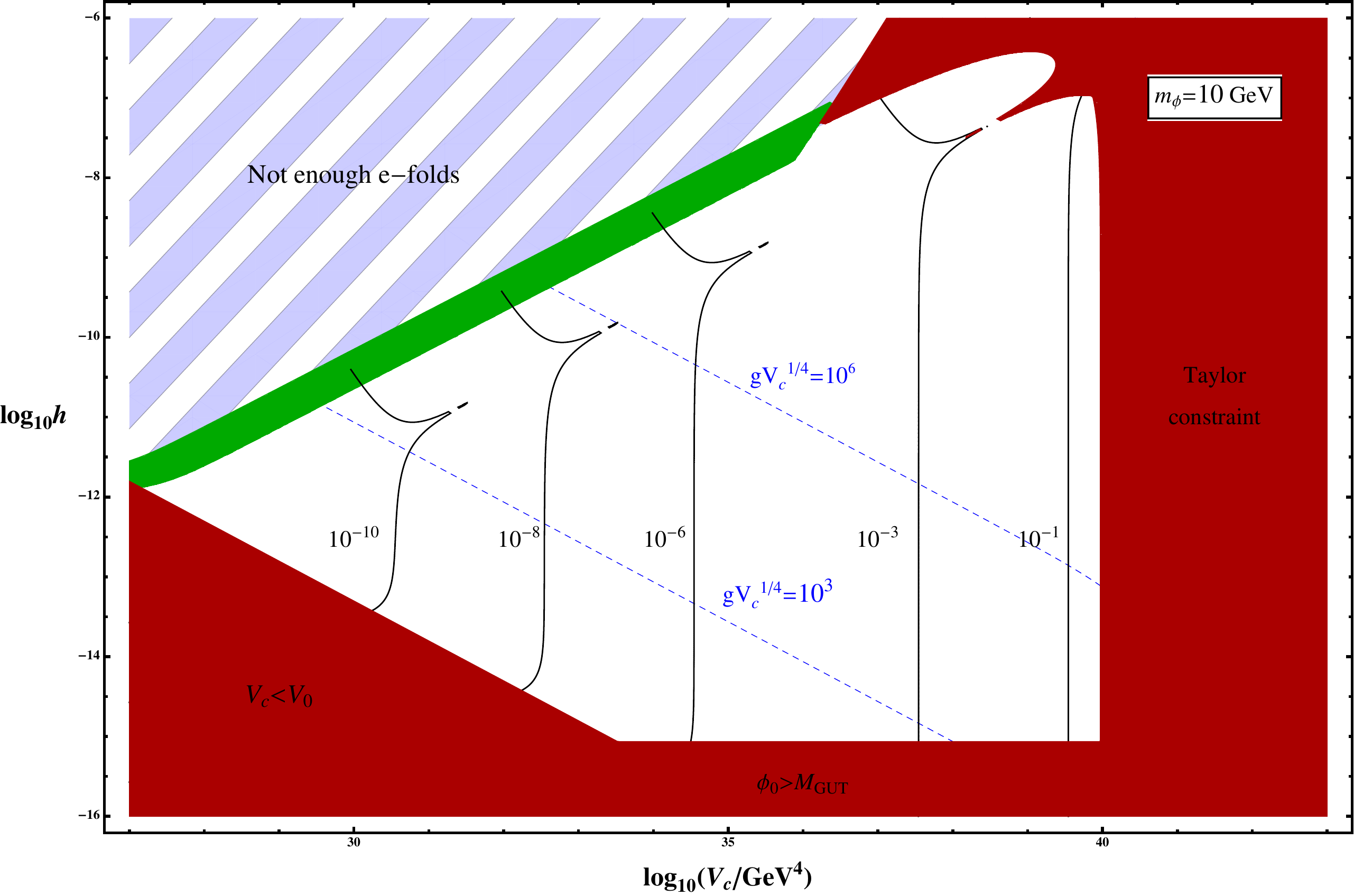}
\caption{Regions of parameter space for the potential in eq.~\eqref{hybridpotential} for the choice $m_\phi(\phi_{0})=10$ GeV. The solid red region is inaccessible to our analysis, for the reasons indicated by the text and explained more fully in section~\ref{section:hybrid}. In the blue striped region, the slow-roll parameters become large and end inflation with number of e-folds $\mathcal{N}<\mathcal{N}_\mathrm{pivot}-5$. The diagonal green contour indicates the region where the number of e-folds of slow-roll inflation is within $\pm5$ of $\mathcal{N}_\mathrm{pivot}$ and no hybrid mechanism is needed to end inflation. In the rest of the plot the WMAP constraints can be satisfied by the choice $\beta=\beta_\mathrm{CMB}$ when the $S$ field ends slow-roll inflation through the hybrid mechanism. Contour lines of $\beta_\mathrm{CMB}$ are shown in black (solid) and those of $gV_c^{1/4}$ are shown in blue (dashed). For other values of $m_\phi$, a qualitatively similar figure is obtained (with appropriate scaling of the axes).}
\label{figure:Vc-h10}
\end{center}
\end{figure*}

Let us consider the case depicted in figure~\ref{figure:Vc-h10},
which shows the $h-V_c$ plane, where we have taken the inflaton mass to be $m_\phi(\phi_{0})=10$ GeV. We impose the following constraints in deciding which regions of the $h-V_c$ plane are accessible to our theory: the field value at the end of inflation, $\phi_e$, should be such that the constraints of eqs.~\eqref{Taylorcondition1} and \eqref{Taylorcondition2} are satisfied and our inflection point analysis is valid; the VEV $\phi_0$ should not exceed the GUT scale; and $V_c>V_0$ in order for the hybrid mechanism to be able to bring inflation to an end. The solid red contour around the boundary to the bottom and right shows the regions that are excluded by any of these criteria. For some values of $V_c$ and $h$, the choice of $\beta_\mathrm{CMB}$ from WMAP spectrum constraints results in a number of e-folds $\mathcal{N}$ that is too low: this region is indicated by the blue striped contour. Clearly the hybrid mechanism cannot make this region accessible to successful inflation but subsequent periods of inflation could by reducing $\mathcal{N}_\mathrm{pivot}$. We do not consider this issue further in this paper. 

The solid green contour extending diagonally upwards shows the region in which the WMAP power spectrum constraints and the e-fold constraint can be simultaneously satisfied, \emph{without} the need for a hybrid end to inflation (as in figure~\ref{figure:m-h}, we use the conservative criterion  $\left(\vert\mathcal{N}-\mathcal{N}_\mathrm{pivot}\vert\right)_\mathrm{min}\leq5$ when $\beta$ is allowed to vary over the range allowed by the WMAP constraints). That is, in this region we require the coupling to satisfy $g\geq2V_c^{1/4}/\phi_e$ with $\phi_e$ given by eq.~\eqref{phieta}, but it is otherwise unconstrained. In the rest of the $h-V_c$ plane (left white) the WMAP spectrum constraints and the e-fold constraint can be satisfied if the $S$ field brings inflation to an early end through its coupling to the inflaton as described above. This is the region of interest to us.

In the region of interest we plot contour lines of $\vert\beta_\mathrm{CMB}\vert$ and of $gV_c^{1/4}$. The value of $gV_c^{1/4}$ gives an indication of the effective mass of the inflaton at the end of inflation. It can be seen that, for $V_c\sim10^{36}-10^{37}~(\mathrm{GeV})^4$ and Yukawa coupling $h\sim10^{-14}-10^{-15}$, we can obtain values as large as
\beq
\label{ft-0}
\vert\beta_\mathrm{CMB}\vert \sim10^{-3}\,,
\eeq
with the inflaton mass around the TeV scale. This value should be compared with $\vert\beta_\mathrm{CMB}\vert\lesssim10^{-12}$ for the equivalent case without the action of the $S$ field. 

This represents a great reduction in the amount of fine-tuning required in the theory. In fact, a further reduction of the fine-tuning can be achieved, with 
\beq
\vert\beta_\mathrm{CMB}\vert\sim10^{-1}
\eeq
or even larger, though for this case the effective mass of the inflaton becomes significantly larger than the TeV scale. 

By lifting the scale of inflation, the mechanism proposed in this paper broadens the slow-roll region of the potential, thus reducing the problem of the fine-tuning of the initial value of $\phi$. The slow-roll region of the potential, where $\vert\eta\vert<1$, is of width $\Delta\phi\sim\mathcal{O}(\vert\beta\vert\phi_0)$. Without the action of the $S$ field, as $\beta\lesssim10^{-12}$, this region is exceedingly narrow and the initial location of the inflaton on the potential appears to be fine-tuned. However the values of $\vert\beta_\mathrm{CMB}\vert\sim10^{-1}$ obtained with addition of the hybrid mechanism greatly broaden the slow-roll region and simultaneously reduce the fine-tuning of initial conditions of $\phi$. (Refs.~\cite{Allahverdi:2008bt,Kamada:2009hy,Allahverdi:2007wh} discuss a mechanism by which a prior period of false vacuum inflation may resolve the problem of initial conditions without fine-tuning; the broadening of the slow-roll region achieved here will make the mechanism discussed in these studies more easily achievable.)

Note also that $h\sim10^{-14}-10^{-15}$ leads to an acceptable Dirac mass $m_\nu\sim10^{-3}-10^{-4}$ eV for the lightest neutrino species, and that all values of $h\lesssim10^{-12}$ can also be obtained. Therefore the explanation for the small neutrino masses made in Refs.~\cite{Allahverdi:2006cx, Allahverdi:2007wt} is not spoiled by our extension of the model.

\subsection{Motivation from particle physics}

The introduction of the singlet field may look like an \emph{ad hoc} choice, but such a field can be accommodated to generate an effective mass  for the right handed (s)neutrino field
through its VEV via an additional superpotential term: $g{\bf S} {\bf N} {\bf N}$. Such a singlet field can naturally occur within the NMSSM (next to Minimal Supersymmetric Standard Model)~\cite{Maniatis:2009re}, where the same scalar $S$ could be responsible for generating an effective $\mu$-term, $\kappa {\bf SH_{u}H_{d}}$, where $\kappa \langle S\rangle \sim100$~GeV. The superpotential terms are
\beq
W=h{\bf NH_{u}L}+g{\bf SNN}+W_{NMSSM}\,.
\eeq
The required vacuum energy, $V_{c}$, can be obtained when the singlet field is settled near its local minimum, $\langle S\rangle \approx 0$, during inflation by virtue of its coupling $g{\bf SNN}$. Note that the inflaton VEV near inflection point is quite large, $\phi_{0}\sim \tilde N\sim 10^{14}$~GeV. The VEV induces an effective mass for $S$. For instance, for the values of parameter which leads to eq.~(\ref{ft-0}), the coupling $g\sim 10^{-6}$, and $m_{S}\sim g\langle \tilde N\rangle \sim 10^{8}$~GeV, which is much larger than the Hubble expansion rate during inflation. For $V_{c}\sim 10^{37}~({\rm GeV})^{4}$, $H_\mathrm{inf}\sim 10$~GeV. Thus the $S$ field settles down in its local minimum 
within one Hubble time during inflation. The value of $g\sim 10^{-6}$ is also adequate to 
generate the right handed (s)neutrino masses after inflation within an expected range: $gV_{c}^{1/4}\sim 10^{3}$ GeV. Finally the phase transition happens when $\phi\rightarrow \phi_{c}$, eq.~(\ref{newphie}). 

For larger values of $m_\phi$ during inflation, achieving a similar increase in $\vert\beta_\mathrm{CMB}\vert$ requires the correction to the inflaton mass after inflation to be greater than $\sim\mathcal{O}(\mathrm{1~TeV})$. This is acceptable in principle, though requiring the mass to be much larger than the TeV scale is somewhat undesirable, and offsets the reduction of the fine-tuning of the inflation potential. 

For the above parameters, the reheat temperature of the universe is sufficiently high to generate the required thermal dark matter abundance. One particularly nice candidate is the lightest right handed sneutrino~\cite{Allahverdi:2007wt,Lee:2007mt,Arina:2007tm,Asaka:2006fs}. However further investigation of how all the MSSM degrees of freedom achieve thermal equilibrium is needed, which we leave for future studies. A preliminary result suggests that the inflaton being gauge invariant generates VEV dependent masses for the $SU(2)$ gauge bosons which delay the process of thermalization and avoid overproduction of thermal gravitinos~\cite{Allahverdi:2005mz,Allahverdi:2008pf,Allahverdi:2007zz}.

\section{Conclusions}
\label{section:conclusion}

We have discussed the very generic renormalizable inflationary potential introduced in eq.~\eqref{genericpotential} and have analyzed the conditions required for this potential to generate a power spectrum of density perturbations compatible with the latest WMAP 7-year constraints. We have also presented a particular particle physics model for the potential \eqref{potential}, where the origin of the inflaton is the gauge-invariant flat direction introduced in eq.~\eqref{superpotential}, with a minimal extension of the standard model gauge group. We obtained the constraints on the parameter space for this model from cosmology, which are depicted in figure~\ref{figure:m-h}. The allowed region of parameter space also allows an explanation for the small neutrino mass. 

The potential \eqref{genericpotential} (or equivalently \eqref{potential}) can be realised in many different theoretical models. As long as the theory allows for rapid transfer of inflaton energy into radiation at the end of inflation, the constraints on parameter space presented in figure~\ref{figure:m-h} will apply. Therefore our results are quite general.

Although in general the soft SUSY-breaking terms in this potential must be highly tuned against each other, we have shown that this is primarily a result of the requirement to produce the correct number of e-folds of inflation, rather than a constraint imposed by the observed power spectrum. With this perspective, we have presented a simple extension of the model to include a new scalar field which brings an end to inflation through the hybrid mechanism. We show this reduces the fine-tuning to very manageable levels, even achieving $\vert\beta\vert\sim10^{-1}$. This result is significant as the required fine-tuning was one of the main objections to the original model. Finally, we argue that this hybrid extension of the model also reduces any need for fine-tuning of the initial value of the inflaton field, which also makes the hybrid extension more attractive.

\acknowledgments

We would like to thank Rhys Davies, Matthew McCullough and Christopher McCabe for stimulating discussions and helpful comments. SH is supported by the Academy of Finland grant 131454. SN is supported by the ORS scheme and the Clarendon Fund from the University of Oxford and Merton College, Oxford.

\end{document}